\documentclass[cits]{PoS}
\usepackage{epsfig}

\title{
{\small DESY 11--258}  \\
{\small WUB/11-28}  \\[2.5cm]
Theoretical improvements for luminosity monitoring at low energies}

\ShortTitle{Theoretical improvements for luminosity monitoring at low energies}

\author{\speaker{Janusz Gluza}, Micha{\l} Gunia \\
Institute of Physics,        University of Silesia, 40-007 Katowice, Poland\\
        E-mail: \email{janusz.gluza@us.edu.pl}}
\author{{Tord Riemann}\\
Deutsches Elektronen-Synchrotron, DESY, Platanenallee
    6, 15738 Zeuthen, Germany }
 
  \author{{Ma{\l}gorzata Worek}\\
 Fachbereich C Physik, Bergische Universit\"{a}t Wuppertal, \\
   Gaussstr. 20,
  D-42097 Wuppertal, Germany
 }

\abstract{A comparison of theoretical results on NNLO leptonic
and hadronic corrections to Bhabha scattering with {the Monte Carlo} generator
BabaYaga@NLO used at meson factories is given.  
Complete NLO virtual corrections to the $e^+e^- \to \mu^+ \mu^- \gamma$ process are discussed.
}

\FullConference{ 10th International Symposium on Radiative Corrections (Applications of Quantum Field Theory to Phenomenology) - Radcor2011\\
September 26-30, 2011\\
Mamallapuram, India}

\begin{document}


\newcommand{\ssL}{{\scriptscriptstyle{L}}}
\newcommand{\ssN}{{\scriptscriptstyle{N}}}
\newcommand{\ssO}{{\scriptscriptstyle{O}}}
\newcommand{\ssQ}{{\scriptscriptstyle{Q}}}
\newcommand{\ssE}{{\scriptscriptstyle{E}}}
\newcommand{\ssD}{{\scriptscriptstyle{D}}}
\newcommand{\ssB}{{\scriptscriptstyle{B}}}
\newcommand{\ssY}{{\scriptscriptstyle{Y}}}

\newcommand{\be}{\begin{equation}}
\newcommand{\ee}{\end{equation}}
\newcommand{\bea}{\begin{eqnarray}}
\newcommand{\eea}{\end{eqnarray}}
\newcommand{\non}{\nonumber}
\newcommand{\nl}{\nonumber \\}

\def \litwo {{\rm{Li_2}}}

\section{Introduction}
This Radcor symposium is devoted mainly to the high energy, LHC physics. However, low energy physics remains an active area of research. There are a few 
on-going  low energy $e^+e^-$ experiments like  the $\Phi$ factory DA$\Phi$NE (with the KLOE detector)
the $B$ factories PEP-II and KEK (with the BaBar detector at SLAC for which the accumulated data are still analysed and the Belle detector at KEK),
the charm/$\tau$ factory  BEPC II (detector BES III).
For them, to operate properly, a
precise calculation of higher order corrections $\sigma_{theory}$ for the process of Bhabha scattering
($e^+e^- \rightarrow e^+e^-$) is necessary. Then the collider's  luminosity $\int L_{tot}  dt=
\frac{N_{exp}}{\sigma_{theory}}$ can be determined with high accuracy. Here $N_{exp}$   stands for a number of
detected events, $\sigma$ is the theoretical cross section of the chosen reference process. If we know the
luminosity accurately in a low energy region, then  also the 
 low energy hadron cross sections in $e^+e^-$ annihilation processes $\sigma_{had} =
\frac{N_{had}}{L_{tot}}=\sigma_{theory} \frac{N_{had}}{N_{exp}}$ can be inspected ($N_{had}$ is the number of
measured hadronic events). What follows is that $\sigma_{theory}$ serves as the calibration parameter in
measurements.\footnote{For more details on possible luminosity options at low energies, see the Introduction
in \cite{Actis:2010gg}.} 
Two loop virtual QED\footnote{Recently the dominant logarithmic 2-loop electroweak corrections, important at ILC energies, have been computed 
in \cite{Penin:2011aa}.} corrections to the Bhabha scattering are already known \cite{Glover:2001ev,Bonciani:2003cj0,Czakon:2004wm,penin:2005kf,Bonciani:2004qt,Bonciani:2004gi,Czakon:2006pa,Actis:2007gi,Becher:2007cu,Bonciani:2007eh,Actis:2007pn,
Actis:2007fs,Actis:2008br,Kuhn:2008zs}, however, so far their effects have not been applied to realistic physical situations (theoretical groups made comparisons only among virtual corrections, switching off the soft-photon cut-off parameter). 
 For unresolved photons, fermion pairs, or hadrons at NNLO (real emission),
\begin{eqnarray}\label{eq-nnlog}
 e^+e^- &\to& e^+e^-(\gamma,\gamma\gamma),~~~  e^+e^-(e^+e^-),  ~~~ e^+e^-(f^+f^
-),  ~~~ e^+e^-(\mathit{hadrons}) ,
\end{eqnarray}
we use the Fortran packages \textsc{Helac--Phegas} (fermion pairs) 
 \cite{Kanaki:2000ey,Papadopoulos:2000tt,Papadopoulos:2005ky,Cafarella:2007pc}, the MC event generator
\textsc{Ekhara} \cite{Czyz:2010sp,Czyz:2006dm,Czyz:2005ab,Czyz:2003gb} (pion pairs), and the Fortran program
\textsc{Bhagen-1Ph-VAC} (photons emission, M.Gunia, H.Czy\.z, unpublished, based
on the  generator  \textsc{Bhagen-1Ph} \cite{Caffo:1996mi}). The complete results are compared with the
 BabaYaga~\cite{CarloniCalame:2000pz,CarloniCalame:2001ny,CarloniCalame:2003yt} MC generator in its recent version \textsc{BabaYaga@NLO}   ~\cite{Balossini:2006wc}.
 
On the other hand,  the muon pair production with real photon emission $e^+e^- \to \mu^+ \mu^- \gamma$ 
 is an important
background and normalization reaction in the measurement of the pion form-factor:
$
    R_{exp} = \frac{\sigma(e^+ e^- \rightarrow \pi^+ \pi^- \gamma)}{\sigma(e^+ e^- \rightarrow \mu^+ \mu^- \gamma)}.
$ 
We report on a progress in the determination of complete virtual corrections.

\section{NNLO Bhabha scattering at low energies}

In \cite{CarloniCalame:2011zq} it has been shown that for assumed realistic event selections the total
NNLO massive corrections are relevant for precision luminosity  measurements with $10^{-3}$ accuracy. Second,
the maximum observed difference between \textsc{BabaYaga@NLO}  and exact massive NNLO corrections is at the level of  0.07\%.
When cuts are varied the sum of the missing pieces
can reach 0.1\%, but for very tight acollinearity cuts only. 
The electron pair contribution is the largely dominant part of the correction. The muon pair and hadronic corrections are the next-to-important effects and quantitatively on the same grounds. The tau pair contribution is negligible for the 
energies of meson factories. Since the work \cite{CarloniCalame:2011zq} has already been reported at two
conferences by M.~Gunia ("Matter To The Deepest", Ustro\'n 2011, \cite{guniaustron}) and G.~Montagna
("PHIPSI11", Novosibirsk 2011, \cite{CarloniCalame:2011aa}), we focus here on a part of results which has not
been discussed there: main aspects of hadronic contributions and narrow resonances, for a complete discussion,
see \cite{CarloniCalame:2011zq}.  
 
Specific problems are caused by narrow resonances like $J/\psi, \psi(2S),\ldots$. Typically, they are very
narrow (at the keV level) and beam spread effetcs (at the MeV level) may cause that their final effects are
diffused. This problem should be study. Here we show that their effects at fixed center of mass energy is
dominating. 

Narrow resonances with mass $M_{\mathit res}$ and partial width $\Gamma^{e^+e^-}_{\mathit res}$ can be described approximately by the ansatz
\begin{equation}
R_{\mathit res}(z)= \frac{9 \pi}{\alpha^2} M_{\mathit res} \Gamma^{e^+ e^-}_{\mathit res}
\delta(z-M^2_{\mathit res}).\, 
\label{eq-reson}
\end{equation}
Based on this, their contributions to the NNLO Bhabha process can be 
 derived from the general formulae of \cite{Actis:2008br}.
We discuss here as an example
 the contribution from the "rest" (genuine irreducible 2-loop corrections, as defined in \cite{Actis:2008br}).
According to Eq. (87) of \cite{Actis:2008br} it reads: 
\begin{equation}
\frac{d {\sigma}_{\rm rest}}{d \Omega}=
\frac{9 \alpha^2}{\pi~s} ~\frac{\Gamma^{e^+ e^-}_{\rm res}}{M_{\rm res}}
\left\{
\frac{F_1(M^2_{\rm res})}{t-M^2_{\rm res}}
+ \frac{1}{s-M^2_{\rm res}} \left[ F_2(M_{\rm res}^2)
+ F_3(M_{\rm res}^2) \ln \left| 1-\frac{M_{\rm res}^2 }{s}\right|
\right]
\right\}.
\label{bhabha-res}
\end{equation} 
 
Let us note that Eq. \ref{bhabha-res} becomes invalid when the center of mass energy comes too close to the position of a resonance, 
i.e. if $(s-M^2_{\rm res}) \lesssim \Gamma^{e^+ e^-}_{\rm res}M_{\rm res}$. In the numerical examples, Table~\ref{table-narrow-res}, this is not the case.

\begin{table}[h]
\setlength{\tabcolsep}{0.3pc}
\caption[]{\emph{
Soft+virtual NNLO contributions $\sigma_{\rm{rest,res}}^{\rm{\ssN\ssN\ssL\ssO}}$ from narrow resonances (n.r.)
defined by Eq.~(\ref{bhabha-res}) for the Bhabha
process  with
$\omega/E_{\rm beam}=10^{-4}$ (in nb).
The narrow resonance located closest to the center of mass energy of the given collider is
included (first column, ${\rm res}$) and excluded (second column, ${\rm res'}$). The third column contains the
Born cross section  $\sigma_B$.
}}
\label{table-narrow-res}
\begin{center}
\begin{tabular}{|lc|l|c|c|}
\hline
 & $\sqrt{s}$ &$\sigma_{\rm{rest,res}}^{\rm{\ssN\ssN\ssL\ssO}}$  & $\sigma_{\rm{rest,res'}}^{\rm{\ssN\ssN\ssL\ssO}}$ &  $\sigma_B$ 
\\
\hline
KLOE   & 1.020 & [all  n.r.]   &  [n.r. without J$\slash \psi$(1S)]
&       \\
       &       &  -0.04538  &   -0.0096      & 529.5  \\\hline
BES    & 3.097 & [all n.r.]    & [n.r. without J$\slash \psi$(1S)]
&        \\
       &       &  {\bf 228.08}         &   {\bf -0.0258}        &  {\bf 14.75}  \\\hline
BES    & 3.650 &  [all n.r.]  & [n.r. without $ \psi$(2S)]     &
\\
       &       &   -0.1907        &    -0.023668        &  123.94  \\\hline
BES    & 3.686 &  [all n.r.]  & [n.r. without $ \psi$(2S)]       &
\\
       &       &   {\bf -62.537}    &     {\bf -0.0254}       &  {\bf 121.53}  \\\hline
BaBar & 10.56 &  [all n.r.]  & [n.r. without $\Upsilon$(4S)]      &
\\
       &       &  -0.0163       &  -0.01438          &  6.744 \\\hline
Belle  & 10.58 &  [all n.r.]  &[n.r. without $\Upsilon$(4S)]       &
\\
       &       &  0.04393   &    -0.0137        &    6.331  \\
\hline
\end{tabular}
\end{center}
\end{table}

 In Table~\ref{table-narrow-res} we show numerical results based on
Eq. (\ref{bhabha-res}).  
We can see that the contributions from narrow resonances dominate the NNLO Bhabha
correction for BES running at $J/\psi$ and $\psi(2S)$ energies.
For the remaining cases narrow resonances contribute below the per mille level
when compared to the Born cross section $\sigma_B$. For those cases results can be found in \cite{CarloniCalame:2011zq}.

We conclude that for experiments performed on top of a narrow
resonance, this resonance cannot be treated as a mere correction and more detailed studies have to be performed.
  These should include the examination of finite width effects, beam spread effects, the estimation of NNNLO
corrections,
  and the accuracy of the vacuum polarisation insertions in a close
  vicinity of these resonances.

\section{$e^+e^-$ $\to$ $\mu^+\mu^-\gamma$ at low energies}

The $e^+e^-$ $\to$ $\mu^+\mu^-\gamma$ process  is an  ideal benchmark process to test for massive tensor reductions at the 1-loop level. 
It has (i) two different masses (and thus large difference of scales, up to 7 orders in magnitude);
(ii) quasi-collinear regions (due to small electron mass); (iii) small number of diagrams.

\begin{figure}[!h] 
\centering
        \includegraphics[scale=0.4]{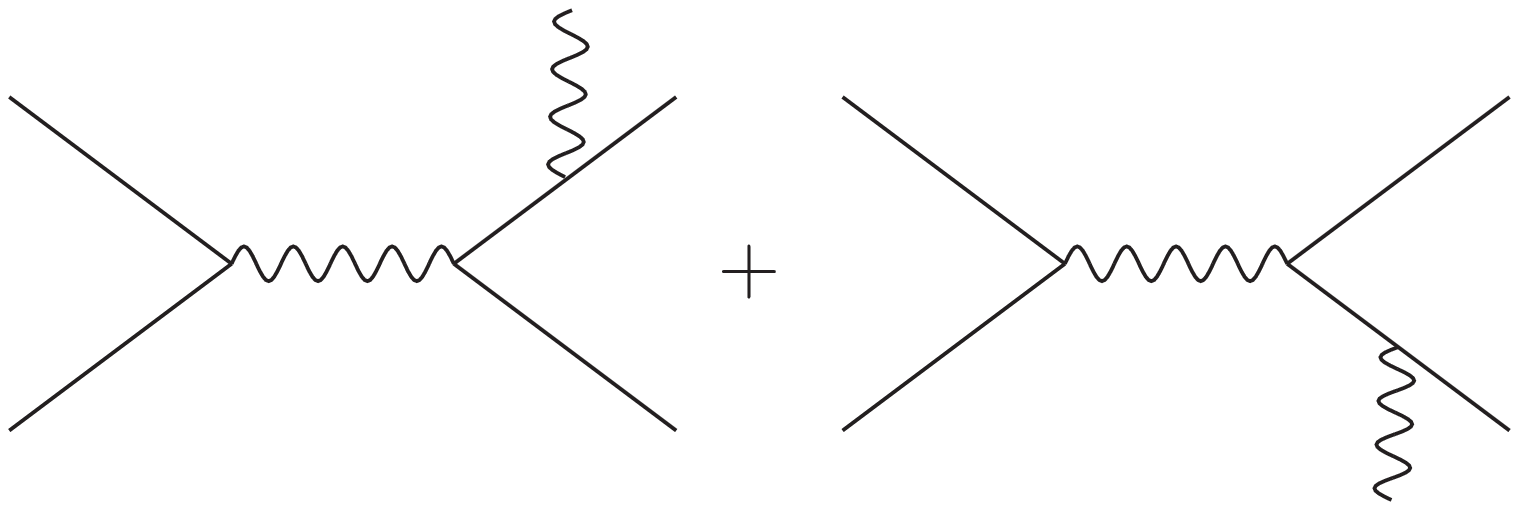}\\
        \includegraphics[scale=0.4]{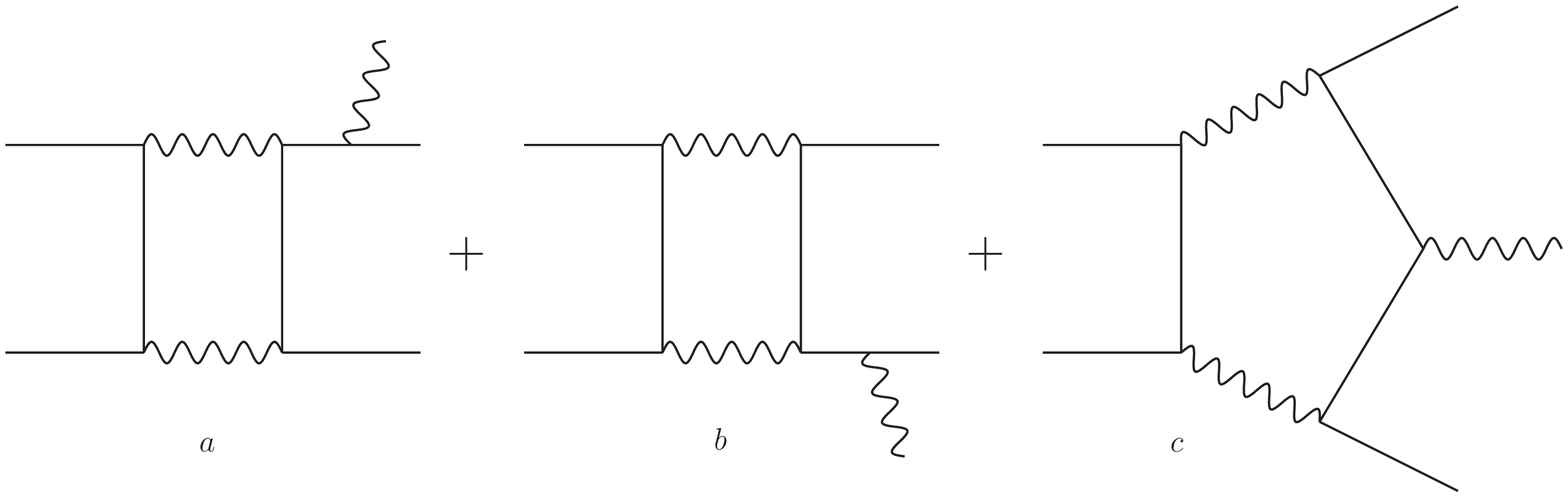}
  \caption{A set of diagrams needed for final state radiation (FSR) gauge invariance of interferences between
tree diagrams (upper picture) and four and five point one-loop integrals (below). Here the diagrams are
limited to FSR cases, the same property holds if initial state radiation (ISR) amplitudes are present.}
  \label{gauge}
\end{figure} 

\begin{figure}[tb] 
\begin{center}
        \includegraphics[scale=0.25,angle=270]{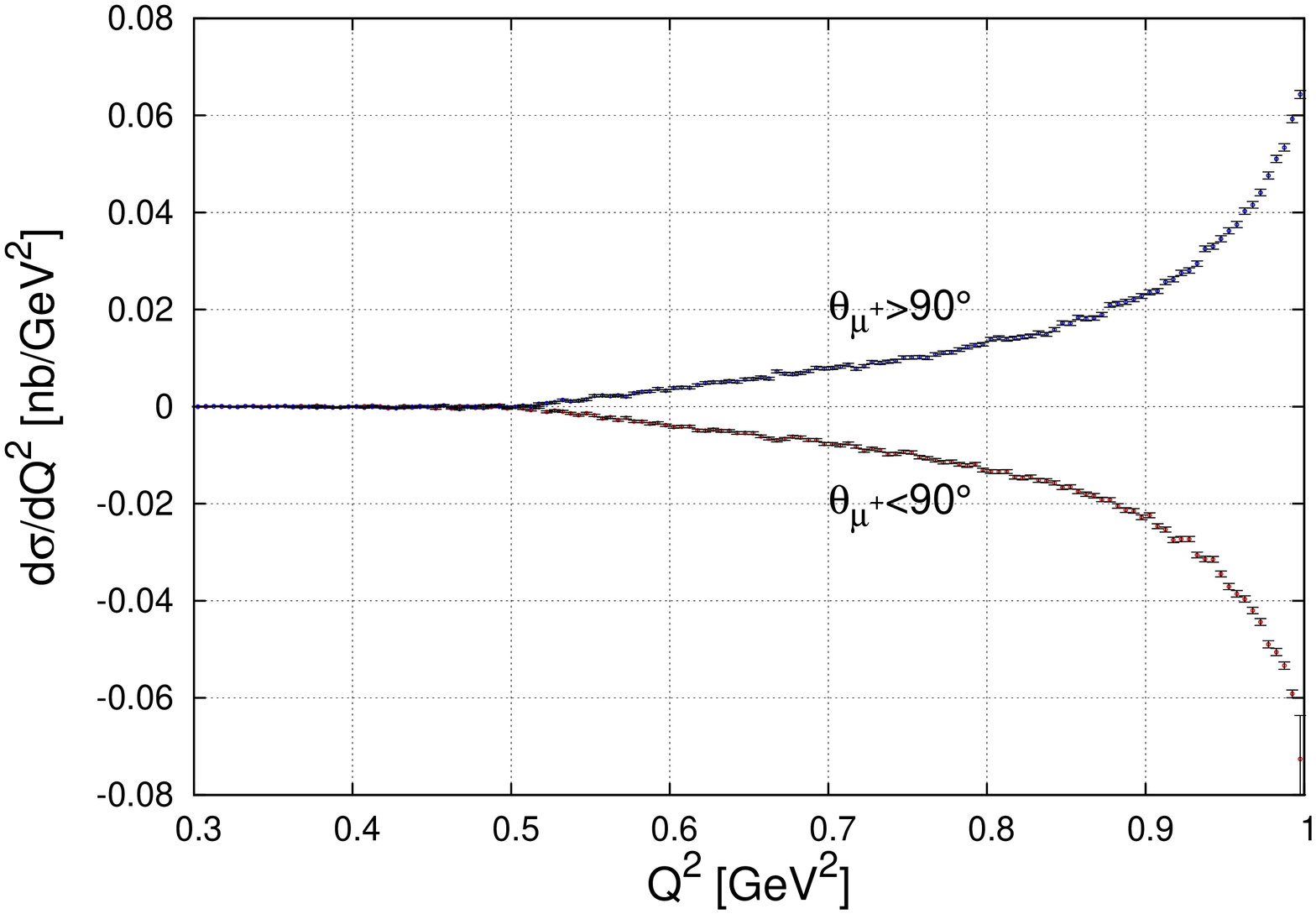}
\end{center}
\caption{Muon pair distributions including 5-point functions at KLOE. $2.5 \times 10^6$ events have been generated.
Looptools and FF packages have been used.}
\label{kloe1}
\end{figure}

\begin{figure}[tb] 
\begin{center}
   \includegraphics[scale=0.25,angle=270]{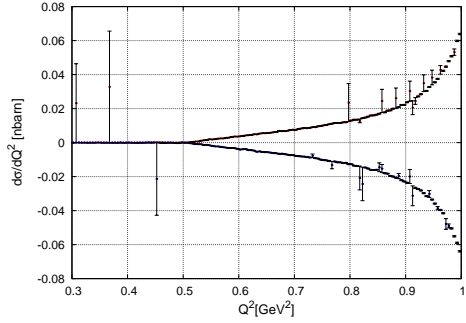} %
\end{center}
\caption{Muon pair distributions including 5-point functions at KLOE. $10^9$ events have been generated.
Looptools and FF packages have been used.}
\label{kloe2}
\end{figure}

\begin{figure}[!h]%
      \centering
       \includegraphics[scale=0.5,angle=0]{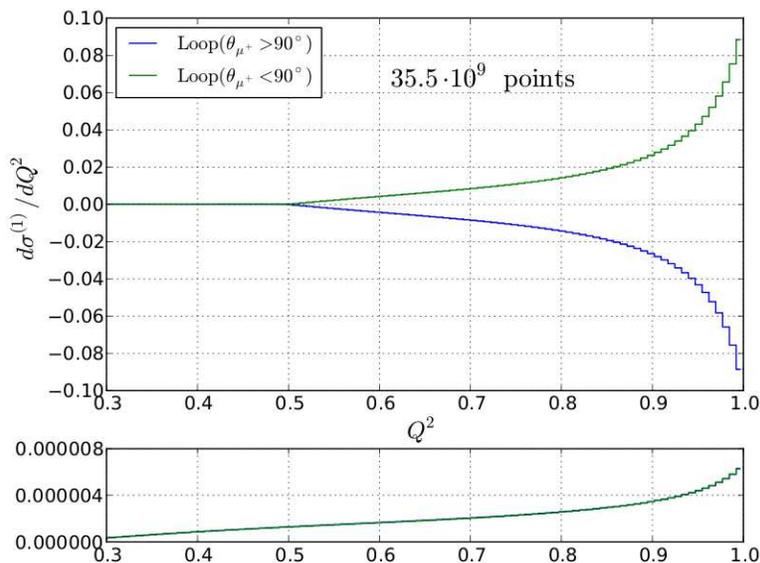} %
\caption{Muon pair distributions including 5-point functions at KLOE. Bottom: absolute error estimate. Approximately $4\cdot 10^{10}$ events has been generated. PJFry package has been used. Taken from \cite{phdyundin}.}
\label{pjfrykloe}
  \end{figure}

There are many tools which give a possibility to obtain automatically cross sections at one-loop level,
however, there is still ongoing progress as new demands on precision of calculations require a more and more
precise numerical analysis. Here we give an example of such a progress in the calculation of effects including
5-point functions. In Fig.\ref{gauge} a class of diagrams is shown needed for building gauge invariant
amplitudes. 
Let us define  the relative accuracy
    \begin{equation}\label{eq:accuracy}
        A=\max  \frac
        {\sum_{i=a,b,c} \Re(M_{\rm loop}^i M_{\rm tree}^\dagger)}
        {\min \Re(M_{\rm loop}^i M_{\rm tree}^\dagger)}
    \end{equation}
    where $i$ stands for (a),(b),(c) in Fig.\ref{gauge}.
For such a  test parameter, with the package \texttt{LoopTools/FF} \cite{Hahn:1998yk,vanOldenborgh:1990yc},
 $2.5 \times 10^6$ events have been generated. For Fortran double 
precision, an accuracy at the level $A=10^{-2}$ has been obtained, and for quadruple precision (with an Intel
Fortran compiler), it is $A=10^{-12}$. 
    It is clear that using quadruple precision, a reasonable level of accuracy has been obtained.
    In Fig.\ref{kloe1} a distribution of muon pairs in $e^+e^-$ $\to$ $\mu^+\mu^-\gamma$ is shown at KLOE
kinematics.\footnote{Similar plots for the same number of generated events but different distributions can
be found in \cite{Kajda:2009aa}).} Parameters as defined in Table~\ref{kloecuts} have been used. The results
in Figs.\ref{kloe1} and \ref{pjfrykloe} estimate only the virtual contributions.
However, with an increasing number of generated events some instabilities appear, see Fig.\ref{kloe2}.
Since version 2.2, \texttt{LoopTools} allows to choose in which decomposition five point integrals should be
calculated. Implementations of both the Passarino-Veltman \cite{Passarino:1978jh} and the Denner/Dittmaier
\cite{Denner:2002ii,Denner:2005nn} schemes exist.
The instability problems are due to the appearance of inverse Gram determinants, and their
 solution is achieved with the alternative reduction scheme 
\cite{Diakonidis:2008ij,Diakonidis:2009fx,Fleischer:2010sq} and its  implementation in 
the package  \texttt{PJFry}.\footnote{For the evaluation of the scalar integrals we use QCDloop/FF
\cite{Ellis:2007qk}. For further references and some basic comparisons see
\cite{Fleischer:2011bi,vanOldenborgh:1990yc}.}

\begin{table}[!h]
\centering
\begin{tabular}{|c|c|c|c|c|}
\hline 
         $E_{CMS}$ & $E_{\gamma,\rm{min}}$ & $\theta_\gamma$                                         & $Q^2$                     
           & $\theta_{\mu^\pm}$  \\ \hline
            1.02 GeV   & 0.02 GeV              & $0^{\circ}$--$15^{\circ}$, $165^{\circ}$--$180^{\circ}$ & $0.25$--$1.06$ $\rm{GeV}^2$ 
            & $50^{\circ}$--$130^{\circ}$ \\ \hline
\end{tabular}
\caption{Phase-space cuts for KLOE settings used in calculations.
 $Q^2$ is the invariant mass squared of the muon pair.}
 \label{kloecuts}
\end{table}

Using these new reductions, in Fig.\ref{pjfrykloe} the same stability of distributions as in
Figs.\ref{kloe1},\ref{kloe2} is obtained.
The results are completely stable and well controlled. E.g. the leading  inverse Gram determinants
$|G^{(5)}|$ are eliminated in the reduction and 
small inverse Gram determinants $|G^{(4)}|$ are avoided using asymptotic expansion.

\section{Summary}
Low energy physics is an important field of activity. Let us mention  the measurement of $(g-2)_{\mu}$  and
its theoretical calculation,\footnote{See the contribution by R.~Szafron, these proceedings.} but also low
energy hadron physics, e.g. the determination of form factors. In this context, an important quantity is the
pion form factor, see e.g. \cite{Hoefer:2001mx}. In order to describe it properly, experimental data are
needed, and for that the process $e^+ e^- \rightarrow \mu^+ \mu^- \gamma$ serves as  normalization reaction.
We have shown some details on progress in the calculation of this process at the NLO level, including 5-point
diagrams. The progress has been possible due to a stable treatment of tensor reductions.
Similarly, theoretical progress in calculations of NNLO effects in Bhabha scattering convenience us that the
existing generator
BabaYaga@NLO  is a reliable tool used for luminosity determination at meson factories. Here we have signaled
only that some additional work on the understanding of narrow resonance contributions to precise Bhabha
scattering studies would be welcome. 

\acknowledgments{JG and TR would like to thank Organizers of the Radcor symposium: D. Indumathi, Prakash
Mathews, Andreas Nyffeler and V. Ravindran for their warm hospitality and perfectly prepared conference.
  We would like to thank: C.C.~Calame, G.~Montagna, O.~Nicrosini, F.~Piccinini (the BabaYaga group) and H.~Czy\.z, J.~Fleischer, V.~Yundin for nice collaborations. 
 This work originated from the activity of the "Working Group on Radiative
 Corrections and Monte Carlo Generators for Low Energies" 
 [\href{http://www.lnf.infn.it/wg/sighad/}{http://www.lnf.infn.it/wg/sighad/}].
MW was supported  by the Initiative and Networking Fund of the 
Helmholtz Association, contract HA-101 ("Physics at the Terascale").
This work is also supported in part by Sonderforschungsbereich/Trans\-re\-gio SFB/TRR 9 of DFG
"Com\-pu\-ter\-ge\-st\"utz\-te Theoretische Teil\-chen\-phy\-sik", 
European Initial Training Network LHCPHENOnet PITN-GA-2010-264564 
and by
Polish Ministry of Science and Higher Education
   from budget for science for 2010-2013 under grant number N N202 102638.}

 
\end{document}